\documentclass[12pt]{ronbun}

\usepackage{color,amsmath,amssymb,graphicx}

\usepackage{cite}
\usepackage{bm}
\usepackage{dcolumn}

\newcommand{\s}{\scriptscriptstyle}

\newcommand{\nn}{\nonumber}
\newcommand{\e}{{\rm e}}

\newcommand{\del}{\delta}

\newcommand{\al}{\alpha}

\newcommand{\ba}{\begin{eqnarray*}}
\newcommand{\ea}{\end{eqnarray*}}

\newcommand{\I}{\bf 1}

\newcommand{\KUCPlogo}{\hbox{\lower 1.4ex\hbox{
\Huge\boldmath $\cal K$}
\kern -1.15em {\sffamily \bfseries\large\ UCP}}
\kern -4.5em \raise 0.2em\hbox{\lower 1.4ex\hbox{\color{cyan}
\Huge\boldmath $\cal K$}
\kern -1.15em {\color{magenta}\sffamily \bfseries\large\ UCP}
\put(-20,-7){\tiny\it preprint}
}}

\numberwithin{equation}{section}

\begin{document}

\begin{flushright}

\parbox{3.2cm}{
{KUCP-0212 \hfill \\
{\tt hep-th/0206132}}\\
\date
 }
\end{flushright}

\vspace*{0.5cm}

\begin{center}
 \Large\bf 
  BPS Conditions of Supermembrane on the PP-wave
\end{center}

\vspace*{1.0cm}

\centerline{\large Katsuyuki Sugiyama$^{\ast}$ 
and Kentaroh Yoshida$^{\dagger}$}

\begin{center}
$^{\ast}$\emph{Department of Fundamental Sciences, \\
Faculty of Integrated Human Studies, \\
Kyoto University, Kyoto, 606-8501, Japan.} \\
{\tt E-mail:~sugiyama@phys.h.kyoto-u.ac.jp}\\
\vspace{0.2cm}
$^{\dagger}$\emph{Graduate School of Human and Environmental Studies,
\\ Kyoto University, Kyoto 606-8501, Japan.} \\
{\tt E-mail:~yoshida@phys.h.kyoto-u.ac.jp}
\end{center}

\vspace*{1.2cm}

\centerline{\bf Abstract}

We study the BPS conditions in the 
closed supermembranes on the maximally 
supersymmetric pp-wave background. 
In particular, the 1/2 and 1/4 BPS states are discussed in detail. 
Moreover, we comment on the zero-modes in the 
invariant mass formulae of the theory.

\vspace*{2.5cm}

\thispagestyle{empty}
\setcounter{page}{0}

\newpage

\section{Introduction}

The matrix model approach \cite{BFSS} to M-theory seems 
a great success. In a recent progress, the pp-wave backgrounds 
\cite{KG,OP,OP2} and Penrose limits \cite{Penrose,Guven} are 
attractive subjects. 
It has been shown that 
the Green-Schwarz (GS) string theory on the pp-wave is exactly-solvable 
\cite{M,MT,RT}. Also, the matrix model on the eleven dimensional 
maximally supersymmetric pp-wave solution \cite{KG} has been proposed in 
Ref.\,\cite{Malda} and further consideration has been done \cite{DSR}.      

In our previous paper \cite{SY},  we have studied the supermembrane 
\cite{sezgin,bergshoeff,dWHN} on
the maximally supersymmetric pp-wave background, and derived the
supercharges and associated algebra with the central charges by carefully
treating the surface terms by the use of the Dirac bracket procedure.

In this paper, we consider the BPS states from the superalgebra derived
in our previous work. We use triangular decomposition for the
supercharge matrix, and derive the BPS conditions by analyzing
the rank of the supercharge matrix. 
In particular, the 1/2 BPS and 1/4 BPS states 
are investigated in detail. We also find that 
the zero-modes of the membrane variables 
do not appear in the BPS conditions.  

In section 2, the BPS states are studied by investigating the rank of the 
supercharge matrix. In particular, we consider the cases of 1/2 and 1/4
BPS states, concretely. Section 3 is devoted to conclusions and
discussions.

\section{Supercharge Matrix and BPS States}

In our previous work \cite{SY}, we have studied supermembranes 
on the maximally supersymmetric pp-wave and 
derived the expressions of the 
supercharges $Q^+$ and $Q^-$. We calculated 
the associated superalgebra. The results are written as follows:  
\begin{eqnarray}
& & i\left\{\frac{1}{\sqrt{2}}Q_{\al}^-,\,\frac{1}{\sqrt{2}}
(Q^-)_{\beta}^{\s T}\right\}_{\rm\s DB} \;=\; 
 - \del_{\al\beta}\, , \\
& & i\left\{\frac{1}{\sqrt{2}}Q_{\al}^+,\,\frac{1}{\sqrt{2}}
(Q^-)_{\beta}^{\s T}\right\}_{\rm\s DB} \;=\; 
  i\sum_{{\s I}=1}^{3} \left[ \left(P_0^{\s I} + \frac{\mu}{3}
X^{\s I}_0 \gamma_{\s 123}\right)\gamma_{\s I} 
\e^{-\frac{\mu}{3}\gamma_{\s 123}\tau}
\right]_{\al\beta}  \\
& & \qquad  +  i\sum_{{\s I'}=4}^{9} \left[ \left(
P_0^{\s I'} - \frac{\mu}{6}X^{\s I'}_0
\gamma_{\s 123} \right)\gamma_{\s I'} 
\e^{-\frac{\mu}{6}\gamma_{\s 123}\tau}
\right]_{\al\beta} - i\sum_{{\s I,J}=1}^3 
\int \! d^2\sigma\, \partial_{a}S^a_{\s IJ}
\left(\gamma^{\s IJ}\e^{-\frac{\mu}{3}\gamma_{\s 123}\tau}\right)_{\al\beta} 
\nn \\ 
& & \qquad - i\sum_{{\s I',J'}=4}^9 
\int \! d^2\sigma\, \partial_{a}S^a_{\s I'J'}
\left(\gamma^{\s I'J'}\e^{-\frac{\mu}{3}\gamma_{\s 123}\tau}\right)_{\al\beta} 
- 2i\sum_{{\s I}=1}^3\sum_{{\s I'}=4}^9 
\int \! d^2\sigma\, \partial_{a}S^a_{\s II'}
\left(\gamma^{\s II'}\e^{-\frac{\mu}{6}\gamma_{\s 123}\tau}\right)_{\al\beta}
\, , \nn 
\end{eqnarray}
\begin{eqnarray}
& & i\left\{\frac{1}{\sqrt{2}}Q_{\al}^+,\,\frac{1}{\sqrt{2}}
(Q^+)_{\beta}^{\s T}\right\}_{\rm\s DB} \;=\; 
2H \delta_{\alpha\beta}  \\
& & \qquad +  \frac{\mu}{3}\sum_{{\s I,J}=1}^3 M^{\s IJ}_0 
\left(\gamma_{\s IJ}\gamma_{\s 123}\right)_{\al\beta} 
- \frac{\mu}{6}\sum_{{\s I',J'}=4}^9 
M_0^{\s I'J'}
\left(\gamma_{\s I'J'}\gamma_{\s 123}\right)_{\al\beta} \nn \\
& & \qquad - 2 \sum_{{\s I}=1}^3
\int\! d^2\sigma\, \varphi X_{\s I}(\gamma^{\s I})_{\al\beta}
-2\sum_{{\s I'=4}}^9\int\!d^2\sigma\,\varphi X_{\s I'}
\left(\gamma^{{\s I'}}\e^{\frac{\mu}{6}\gamma_{\s 123}\tau}\right)_{\al\beta} 
\nn \\
& & \qquad + 2 \sum_{{\s I}=1}^3
\int\! d^2\sigma\, \partial_a S^{a}_{\s I}  (\gamma^{\s I})_{\al\beta}  
+ 2 \sum_{{\s I'}=4}^9
\int\!d^2\sigma\, \partial_a S^{a}_{\s I'}
 \left(\gamma^{\s I'}\e^{\frac{\mu}{6}\gamma_{\s 123}\tau}
\right)_{\al\beta}
\nn \\       
& & \qquad + 2\sum_{{\s I,J}=1}^3\sum_{{\s I',J'}=4}^9
\int\!d^2\sigma\,
\partial_{a}S^{a}_{\s IJI'J'} \left(\gamma^{\s IJI'J'}\right)_{\al\beta} 
+ 2\!\!\! \sum_{{\s I',J',K',L'}=4}^9\int\!d^2\sigma\,
\partial_{a}S^{a}_{\s I'J'K'L'} \left(\gamma^{\s I'J'K'L'}\right)_{\al\beta} 
\nn \\
& & \qquad  + 2 \sum_{{\s I,J,K}=1}^3\sum_{{\s I'}=4}^9
\int\!d^2\sigma\,
\partial_{a}S^{a}_{\s IJKI'} \left(\gamma^{\s IJKI'}\e^{\frac{\mu}{6}
\gamma_{\s 123}\tau}\right)_{\al\beta}  \nn \\
& & \qquad 
+ 2 \sum_{{\s I}=1}^3\sum_{{\s I',J',K'}=4}^9
\int\!d^2\sigma\,
\partial_{a}S^{a}_{\s II'J'K'}  \left(\gamma^{\s II'J'K'}\e^{\frac{\mu}{6}\gamma_{\s 123}\tau}
\right)_{\al\beta} 
\nn \\
& & \qquad + 2\mu\sum_{{\s J,K}=1}^3\sum_{{\s I',J'}=4}^9
\int\!d^2\sigma\,
\partial_{a}U_{\s JKI'J'}^{a}\left(\gamma^{\s JK}\gamma^{\s I'J'}\right)_{\al\beta} \nn \\ 
& & \qquad 
+ 2\mu \sum_{{\s I'}=4}^9\int\!d^2\sigma\,
\partial_{a}U_{\s I'}^{a}\left(\gamma^{\s I'}\gamma_{\s 123}\e^{\frac{\mu}{6}\gamma_{\s 123}\tau}\right)_{\al\beta} 
\,. \nn
\end{eqnarray}
Here $M^{\s IJ}_0$ and $M^{\s I'J'}_0$ are defined by 
\begin{eqnarray}
& &  M^{\s IJ}_0 \;\equiv\; \int\!d^2\sigma \,
\left(X^{\s I}P^{\s J} - P^{\s I}X^{\s J} - 
\frac{1}{2}S\gamma^{\s IJ}\psi\right)\,, \\
& &  M^{\s I'J'}_0 \;\equiv\; \int\!d^2\sigma\,
\left(X^{\s I'}P^{\s J'} - P^{\s I'}X^{\s J'} - 
\frac{1}{2}S\gamma^{\s I'J'}\psi\right)\,,
\end{eqnarray} 
where $P^r (\equiv w D_{\tau}X^{r}) $ and 
$S_{\al} (\equiv iw \psi^{\s T}_{\al})$ are 
the canonical momenta of $X^r$ and $\psi$,
respectively. 
The zero-modes of $P^r$ and $X^r$ are 
defined by 
\begin{eqnarray}
 P_0^r &\equiv& \int\! d^2\sigma\, w D_{\tau}X^{r}\, , 
\,\,\,X^{r}_0 \;\equiv\; \int\!d^2\sigma\,wX^{r}\,,
\end{eqnarray}
and describe the motion of the membrane's center of mass.
Also, the Hamiltonian $H$ is expressed by  
\begin{eqnarray}
& & H \;=\; 
\int\!d^2\sigma\,w\Bigg[\frac{1}{2}\left(\frac{P^r}{w}\right)^2
+ \frac{1}{4} \{X^r,X^s\}^2
+ \frac{1}{2} \left(\frac{\mu}{3}\right)^2\sum_{{\s I}=1}^{3}(X^{\s I})^2
+ \frac{1}{2} \left(\frac{\mu}{6}\right)^2\sum_{{\s I'}=4}^{9}(X^{\s I'})^2
\nn\\
& &\qquad +\frac{\mu}{6}\sum_{{\s I,J,K}=1}^{3}\epsilon_{\s IJK}
X^{\s K} \{X^{\s I},X^{\s J} \} - w^{-1}\frac{\mu}{4}S\gamma_{\s 123}
\psi -w^{-1}S\gamma_{r}\{X^r,\psi\}\Bigg]\,.
\end{eqnarray}
Other quantities in the above algebra are defined by
\begin{eqnarray}
 S^a_{rs} &\equiv& -\frac{1}{2}\epsilon^{ab}X^{[r}\partial_b X^{s]}\, , \\
 \varphi &\equiv& w \{w^{-1}P^r,\,X^r\} + iw \{\psi^{\s T},\,\psi\}\, , \\
 S^a_r &\equiv& \epsilon^{ab}
\left(w^{-1}X_rP_s\partial_{b}X^s + X_r i\psi^{\s T}\partial_{b}\psi
 + \frac{3}{8}iX^s\partial_b(\psi^{\s T}\gamma_{rs}\psi)\right)\, ,  \\ 
 S^a_{rstu} &\equiv& \frac{i}{48}\epsilon^{ab}X_{[r}\partial_b \left(
\psi^{\s T}\gamma_{stu]}\psi\right)
\, ,\\
 U_{\s JKI'J'}^a &\equiv& - \frac{1}{12}\sum_{I=1}^3
\epsilon_{\s IJK}\epsilon^{ab} X^{\s I'}\partial_{b}(
X^{\s I}X^{\s J'})\,, \\ 
 U_{\s I'}^a &\equiv& - \frac{1}{2}
\epsilon^{ab} X^{\s I'}\partial_{b}
\left[\frac{1}{3}\sum_{{\s I=1}}^3(X^{\s I})^2 
- \frac{1}{6}
\sum_{{\s J'}=4}^9
(X^{\s J'})^2 
\right]\,. 
\end{eqnarray}

In order to study the BPS states in the supermembrane theory on the
pp-wave background, let us construct the supercharge matrix 
with $32\times 32$ components\footnote{Hereafter, we use the expressions 
of the supercharges redefined by rearranging the
factor 1/$\sqrt{2}$ into the overall factor of the fermion $\psi$. } 
\begin{eqnarray}
 i\{Q_{\al},\,Q_{\beta}^{\s T}\}_{\rm\s DB} & \equiv &  
\begin{pmatrix} 
i\{Q^{-}_{\al},\, (Q^{-})_{\beta}^{\s T}\}_{\rm\s DB} &  i\{Q^{-}_{\al},\, 
(Q^{+})_{\beta}^{\s T}\}_{\rm\s DB} \\
i\{Q^+_{\al},\, (Q^{-})_{\beta}^{\s T} \}_{\rm\s DB} &   
i\{Q^{+}_{\al},\, (Q^{+})_{\beta}^{\s T}\}_{\rm\s DB} 
\end{pmatrix}\, .
\end{eqnarray}
By the use of the formula of the triangular decomposition, which can be
applied for an arbitrary matrix, 
\begin{eqnarray}
 \begin{pmatrix}
  A & B \\
  C & D 
 \end{pmatrix}
&=& 
 \begin{pmatrix}
  \I & BD^{-1} \\
  0 & \I 
 \end{pmatrix}
 \begin{pmatrix}
  A - BD^{-1}C & 0 \\
  0 & D  
 \end{pmatrix} 
 \begin{pmatrix}
  \I & 0 \\
  D^{-1}C & \I 
 \end{pmatrix}
\nn \\
&=& 
 \begin{pmatrix} 
  \I & 0 \\
 CA^{-1} & \I 
 \end{pmatrix}
 \begin{pmatrix}
  A & 0 \\
  0 & D - CA^{-1}B
 \end{pmatrix} 
 \begin{pmatrix}
  \I & A^{-1}B \\
  0 & \I 
 \end{pmatrix}
\, ,
\end{eqnarray}
we can decompose the supercharge matrix as follows:
\begin{eqnarray}
 \begin{pmatrix}
 \I & 0 \\
 N_2 & \I 
 \end{pmatrix}
\begin{pmatrix}
\I & 0 \\
0 & \exp\left(-\frac{\mu}{12}\gamma_{\s 123}\tau\right)
\end{pmatrix}
 \begin{pmatrix}
 - \I & 0 \\
  0 &  {\rm m}  
 \end{pmatrix}
\begin{pmatrix}
\I & 0\\
0 & \exp\left(\frac{\mu}{12}\gamma_{\s 123}\tau\right)
\end{pmatrix}
 \begin{pmatrix}
  \I & N_1 \\ 
  0 & \I 
 \end{pmatrix}\, .
\end{eqnarray}
Each block part is a matrix with $16\times 16$ components.
The component matrix ``m'' in the triangular decomposed expression 
is given by  
\begin{eqnarray}
 {\rm m}_{\gamma\delta} 
&=& \Bigg[\,2H - \sum_{r=1}^9(P_0^r)^2 - 2 \sum_{r,s=1}^9 z_{rs}z^{rs} 
+ \frac{2}{3}\mu \sum_{{\s I,J,K}=1}^3\epsilon_{\s IJK}X^{\s I}_0z^{\s JK} \nn \\ & & 
- \left(\frac{\mu}{3}\right)^2\sum_{{\s I}=1}^3(X_0^{\s I})^2 
-\left(\frac{\mu}{6}\right)^2 \sum_{{\s I'}=4}^9(X_0^{\s I'})^2
\Bigg]\delta_{\gamma\delta} + 2\sum_{r=1}^9\left[z_r 
+ 2 \sum_{s=1}^9 P_0^sz_{rs} \right](\gamma^r)_{\gamma\delta} \nn \\
& & -2\sum_{r=1}^9\int\!d^2\sigma\,\varphi X^r (\gamma^r)_{\gamma\delta} 
+ \sum_{r,s,t,u = 1}^9 z_{rs}z_{tu}(\gamma^{rstu})_{\gamma\delta} 
+ 2\sum_{r,s,t,u=1}^9z_{rstu}(\gamma^{rstu})_{\gamma\delta}
\nn  \\ 
& & + 2\mu \sum_{{\s J,K}=1}^3\sum_{{\s I',J'}=4}^9
\left[U_{\s JKI'J'} - \frac{1}{6}
\sum_{{\s I}=1}^3\epsilon_{\s IJK}(X_0^{\s I}z_{\s I'J'} + X_0^{\s I'}
z_{\s IJ'})
\right]
(\gamma_{\s JK}\gamma_{\s I'J'})_{\gamma\delta} \nn \\
& & + 4\mu \sum_{{\s I'}=4}^9\left[\frac{1}{2}U_{\s I'} 
+ \frac{1}{3}\sum_{{\s I}=1}^3
X_0^{\s I}z_{{\s I I'}} 
- \frac{1}{6}\sum_{{\s J'}=4}^9X_0^{\s J'}z_{\s J'I'}
\right](\gamma_{\s I'}\gamma_{\s 123})_{\gamma\delta}\, \nn\\
& & 
+\frac{\mu}{3}\sum_{{\s I,J}=1}^3
\left[M^{\s IJ}_0-(X^{\s I}_0P^{\s J}_0-
X^{\s J}_0P^{\s I}_0)
\right](\gamma_{\s IJ}\gamma_{123})_{\gamma\delta}\nn\\
& &
-\frac{\mu}{6}\sum_{{\s I',J'}=4}^9
\left[M^{\s I'J'}_0-(X^{\s I'}_0P^{\s J'}_0-
X^{\s J'}_0P^{\s I'}_0)
\right](\gamma_{\s I'J'}\gamma_{123})_{\gamma\delta}\,,
\label{m}
\end{eqnarray}
where the central charges $z_{rs}$, $z_r$, $z_{rstu}$,  
$U_{\s JKI'J'}$ and $U_{\s I'}$ are
written as 
\begin{eqnarray}
 z_{rs} &=& \int\!d^2\sigma\,\partial_aS^a_{rs}=
- \frac{1}{2}\int\!d^2\sigma\, w \{ X_r,\, X_s \}\, , \\
 z_{r} &=& \int\!d^2\sigma\,\partial_aS^a_r=
\sum_{s =1}^9 \int\!d^2\sigma\, w\{w^{-1}X_r P_s,\, X^s\}\,,\nn\\
&&\qquad +i\int\!d^2\sigma\,w\{X_r\psi^{\s T},\psi\}
+\frac{3}{8}i\sum_{s=1}^9\int\!d^2\sigma\,w\{X^s,\psi^{\s T}\gamma_{rs}\psi\}
\,,\\ 
z_{rstu} &=& \int\! d^2\sigma\,\partial_a S^a_{rstu} 
= \frac{i}{48}\int\! d^2\sigma\, w \{X_{[r},\psi^{\s T}\gamma_{stu]}\psi\}\,, 
\\  
 U_{\s JKI'J'} &=& - \frac{1}{12} \sum_{{\s I}=1}^3\epsilon_{\s IJK}
\int\!d^2\sigma\, w \{X^{\s I'},\,X^{\s I}X^{\s J'} \}\, , \\ 
 U_{\s I'} &=& - \frac{1}{2}\int\! d^2\sigma\, 
w \left\{X^{\s I'},\, \frac{1}{3}
\sum_{{\s I}=1}^3(X^{\s I})^2 - \frac{1}{6}\sum_{{\s J'}=4}^9 (X^{\s J'})^2
    \right\}\, ,
\end{eqnarray}
and constraint $\varphi$ is given by
\begin{eqnarray}
\varphi =w\{w^{-1}P^r,X^r\}+iw\{\psi^{\s T},\psi\}\,. 
\end{eqnarray}
Also, the component matrices $N_1$ and $N_2$ in the right and left 
triangle matrices are respectively written as 
\begin{eqnarray}
 N_1 &=& - i\sum_{r=1}^9P_0^r \left(\e^{\frac{\mu}{4}\gamma_{\s 123}\tau}\gamma_r\e^{\frac{\mu}{12}\gamma_{\s 123}\tau}\right) - i\sum_{r,s=1}^9 z_{rs}
\left( \e^{\frac{\mu}{4}\gamma_{\s 123}\tau}\gamma^{rs}\e^{\frac{\mu}{12}
\gamma_{\s 123}\tau}  \right) \nn \\
& & + i\frac{\mu}{3}\sum_{{\s I}=1}^3 
X^{\s I}_0\left( \e^{\frac{\mu}{4}\gamma_{\s 123}\tau}\gamma_{\s I}\gamma_{\s 123}\e^{\frac{\mu}{12}
\gamma_{\s 123}\tau}  \right) - i\frac{\mu}{6}\sum_{{\s I'}=4}^9 X^{\s I'}_0
\left( \e^{\frac{\mu}{4}\gamma_{\s 123}\tau}\gamma_{\s I'}
\gamma_{\s 123}\e^{\frac{\mu}{12}\gamma_{\s 123}\tau}  \right)\, ,  \\
 N_2 &=& -i\sum_{r=1}^9 P_0^r 
\left(\e^{-\frac{\mu}{12}\gamma_{\s 123}\tau} \gamma_r 
\e^{- \frac{\mu}{4}\gamma_{\s 123}\tau}\right) + i \sum_{r,s=1}^9 
z_{rs} \left(\e^{-\frac{\mu}{12}\gamma_{\s 123}\tau} \gamma^{rs} 
\e^{- \frac{\mu}{4}\gamma_{\s 123}\tau}\right) \nn \\
& & -i \frac{\mu}{3} \sum_{{\s I}=1}^3 X^{\s I}_0 
 \left(\e^{-\frac{\mu}{12}\gamma_{\s 123}\tau} \gamma_{\s I}\gamma_{\s 123} 
\e^{- \frac{\mu}{4}\gamma_{\s 123}\tau}\right) 
-i  \frac{\mu}{6} \sum_{{\s I'}=4}^9 X^{\s I'}_0 
 \left(\e^{-\frac{\mu}{12}\gamma_{\s 123}\tau} \gamma_{\s I'}\gamma_{\s 123} 
\e^{- \frac{\mu}{4}\gamma_{\s 123}\tau}\right)\,  \nn\\
&=& N_1^{\s T}\,.
\end{eqnarray} 

Here, we should note on the invariant mass $\mathcal{M}$, which plays 
an important role in studying the BPS states in the theory.  
Let us recall that the invariant mass $\mathcal{M}$ of a supermembrane 
is defined by 
\begin{eqnarray}
 \mathcal{M}^2 \; \equiv \;  - P_0^{\hat{\mu}}P_{\hat{\mu}0} \;=\; 
2H - \sum_{r=1}^9P_{0}^rP_0^r, \quad (P_0^{+} =1)\,.
\end{eqnarray}
Thus the first two terms of Eq.\,(\ref{m}) can be replaced with the 
invariant mass $\mathcal{M}^2$ of a supermembrane. 
In particular, if we consider 1/2 BPS states of 
the closed supermembrane in the 
flat space (i.e., the $\mu \rightarrow 0$ limit),  
then BPS condition 
\begin{eqnarray}
 \mathcal{M}^2 \; = \; 2\sum_{r,s =1}^9z_{rs}z^{rs}\, ,
\label{flat}
\end{eqnarray} 
arises from the requirement that the  coefficient of
$\delta_{\gamma\delta}$ 
equals zero. This condition implies that a mass should be proportional 
to its charge in the BPS state. 

Hereafter, we shall restrict ourselves to the case 
of the closed membrane. 

In flat case 
the Hamiltonian contains only the momentum zero-modes $P_0^r$'s and not 
$X^r$s' zero-modes $X_0^r$'s and fermion zero-modes $\psi_0$'s. However, 
the invariant mass $\mathcal{M}^2$ includes no zero-modes since the
momentum zero-modes are subtracted by definition. 

In the pp-wave case, due to the presence of the 
additional terms in the action 
(3-point coupling, boson mass terms and fermion mass term), the Hamiltonian
includes the zero-modes $X_0^r$ and $\psi_0$ in addition to the momentum
zero-mode $P^r_0$ even if we consider closed membranes. In fact, 
we will see that these zero-modes are subtracted and do not appear  
in our considerations for the BPS states. This result might be 
understood if one notes that the $U(1)$ parts decouple from $SU(N)$ parts
even in the pp-wave background as noted in Refs.\,\cite{Malda,DSR}.

From now on, 
we will show that the matrix ``m'' is independent of the zero-modes.  
Let us notice the part in ``m'' proportional to $\delta_{\gamma\delta}$.
First we will decompose the coordinate $X^r$ and momentum $P^r$ 
around their averaged values
\begin{eqnarray}
 & & X^r \;=\; X_0^r + \tilde{X}^r\,, \quad 
X_0^r \;=\; \int\! d^2\sigma\,w X^r\,,  \\
& & P^r \;=\; wP_0^r + \tilde{P}^r\,, \quad 
P_0^r \;=\; \int\! d^2\sigma\, P^r\,. 
\end{eqnarray}
Then the focusing part can be rewritten as
\begin{eqnarray}
\tilde{H} &\equiv& H - \frac{1}{2}\sum_{r=1}^9(P^r_0)^2
+\frac{\mu}{3} \sum_{{\s I,J,K}=1}^3\epsilon_{\s IJK}X^{\s I}_0z^{\s JK}
\nonumber\\
&&\qquad - \frac{1}{2}\left(\frac{\mu}{3}\right)^2
\sum_{{\s I}=1}^3(X^{\s I}_0)^2
- \frac{1}{2}
\left(\frac{\mu}{6}\right)^2\sum_{{\s I'}=4}^9(X^{\s I'}_0)^2 \nonumber\\
&=&  \int\!d^2\sigma\,w\Biggl[
\frac{1}{2}\left(\frac{\tilde{P}^r}{w}\right)^2+
\frac{1}{2}\{\tilde{X}^r,\tilde{X}^s\}^2
+\frac{\mu}{6}\sum_{{\s I,J,K}=1}^3
\epsilon_{\s IJK}\tilde{X}^{\s K}\{\tilde{X}^{\s I},\tilde{X}^{\s J}\}
\nonumber\\
&&\qquad +\frac{1}{2}\left(\frac{\mu}{3}\right)^2
\sum_{{\s I}=1}^3(\tilde{X}^{\s I})^2
+\frac{1}{2}\left(\frac{\mu}{6}\right)^2
\sum_{{\s I'}=4}^9(\tilde{X}^{\s I'})^2\nonumber\\
&&\qquad 
-w^{-1}\frac{\mu}{4}S\gamma_{123}\psi -w^{-1}S
\gamma_r\{\tilde{X}^r
,\psi\}
\Biggr]
\,. 
\end{eqnarray}
The $\tilde{H}$ is interpreted as a Hamiltonian 
where bosonic zero modes $X_0$ and $P_0$ are
subtracted. 
As a result, the BPS condition on the pp-wave corresponding to 
one (\ref{flat}) in the flat case, is given by 
\begin{eqnarray}
 2\tilde{H} \;=\; 2 \sum_{r,s =1}^9z_{rs}z^{rs}\,.  
\end{eqnarray}
Now, we may understand the subtracted Hamiltonian $\tilde{H}$ as 
the improved invariant mass $\tilde{\mathcal{M}}$ defined by 
\begin{eqnarray}
 \tilde{\mathcal{M}}^2 &\equiv& 2\tilde{H}  \nn \\
&=& \mathcal{M}^2 + \frac{2}{3}\mu \sum_{{\s I,J,K}=1}^3
\epsilon_{\s IJK}X^{\s I}_0z^{\s JK}\nonumber\\
& & - \left(\frac{\mu}{3}\right)^2
\sum_{{\s I}=1}^3(X^{\s I}_0)^2
- \left(\frac{\mu}{6}\right)^2\sum_{{\s I'}=4}^9(X^{\s I'}_0)^2\,,
\end{eqnarray} 
where the zero-modes $X_0^r$'s are completely subtracted. 
Therefore, we find that the zero-modes in the first square bracket 
of (\ref{m}) are spurious.  

Next, in the similar way we can subtract the zero modes from the 
$U_{\s JKI'J'}$ and $U_{\s I'}$, and obtain the following expressions 
\begin{eqnarray}
 \tilde{U}_{\s JKI'J'} &\equiv& - \frac{1}{12}\sum_{{\s I}=1}^3 
\epsilon_{\s IJK}\int\! 
d^2\sigma\, w \{\tilde{X}^{\s I'},\,\tilde{X}^{\s I}\tilde{X}^{\s J'}\} \nn \\
&=& U_{\s JKI'J'} - \frac{1}{6}\sum_{{\s I}=1}^3
\epsilon_{\s IJK}(X_0^{\s I}z_{\s I'J'} + X_0^{\s I'}z_{\s IJ'})\, , \\
 \frac{1}{2}\tilde{U}_{\s I'} &\equiv& - \frac{1}{4}\int\!d^2\sigma\, w 
\left\{\tilde{X}^{\s I'},\, \frac{1}{3}\sum_{{\s I}=1}^3(\tilde{X}^{\s I})^2 - \frac{1}{6}
\sum_{{\s J'}=4}^9 (\tilde{X}^{\s J'})^2  \right\} \nn \\
&=& \frac{1}{2}U_{\s I'} + \frac{1}{3}\sum_{{\s I}=1}^3 X_0^{\s I}z_{\s II'} 
- \frac{1}{6}\sum_{{\s J'}=4}^9X_0^{\s J'}z_{\s J'I'}\,.
\end{eqnarray}
Also, the charges $z_r$, $z_{rs}$ and $z_{rstu}$ are rewritten as  
\begin{eqnarray}
 \tilde{z}_r &\equiv& \sum_{s=1}^9\int\! d^2\sigma\, w
\{w^{-1}\tilde{X}^r\tilde{P}_s,\, 
\tilde{X}^s\} \nn\\
&&  +i\int\!d^2\sigma\,w\{\tilde{X}_r\psi^{\s T},\psi\}
+\frac{3}{8}i\sum_{s=1}^9
\int\!d^2\sigma\,w\{\tilde{X}^s,\psi^{\s T}\gamma_{rs}\psi\}\nn\\
&=& z_r + 2 \sum_{s=1}^9
P_0^s z_{rs} - X_0^r \int\! d^2\sigma\, \tilde{\varphi}\,,   \\
 \tilde{\varphi} &\equiv& 
\sum_{r=1}^9 w \{w^{-1}\tilde{P}^r,\, \tilde{X}^r \}
+iw\{\psi^{\s T},\psi\}\,,  \\
\tilde{z}_{rs} &\equiv& - \frac{1}{2} \int\! d^2\sigma\, 
w\{\tilde{X}^r,\, \tilde{X}^s\} \;=\; z_{rs}\,, \\ 
\tilde{z}_{rstu} &\equiv& \frac{i}{48}\int\!d^2\sigma\,w\{\tilde{X}_{[r},\psi^{\s T}\gamma_{stu]}\psi\} \;=\; z_{rstu}\,.
\end{eqnarray}
Thus, it is shown that the zero-modes in 
square brackets containing $z_r$, $U_{\s JKI'J'}$, $U_{\s I'}$ of (\ref{m})
are also spurious. 
Moreover, we can easily show that the other terms in (\ref{m}) do not
contain any bosonic zero-modes, and so 
we obtain the drastically simplified result for the matrix ``m'' as 
\begin{eqnarray}
 {\rm m}_{\gamma\delta} &=& 2\left[
\tilde{H}- \sum_{r,s=1}^9z_{rs}z^{rs}\right]\delta_{\gamma\delta} 
+ 2 \sum_{r=1}^9
\left[
\tilde{z}_r-\int\!d^2\sigma \tilde{\varphi}\tilde{X}^r
\right]
(\gamma^{r})_{\gamma\delta} \nn \\
&& + 
\sum_{r,s,t,u=1}^9 \left[z_{rs}z_{tu} +2 z_{rstu}\right]
(\gamma^{rstu})_{\gamma\delta}
\nn \\
& & + 2\mu \sum_{{\s J,K}=1}^3\sum_{{\s I',J'}=4}^9 \tilde{U}_{\s JKI'J'}
(\gamma_{\s JK I'J'})_{\gamma\delta} + 2 \mu \sum_{{\s I'}=4}^9 
\tilde{U}_{\s I'}(\gamma_{\s I'}\gamma_{\s 123})_{\gamma\delta}\,\nn\\
& & +\frac{\mu}{3}\sum_{{\s I,J}=1}^3
\tilde{M}^{\s IJ}_0
 (\gamma_{\s IJ}\gamma_{123})_{\gamma\delta}
  -\frac{\mu}{6}\sum_{{\s I',J'}=4}^9
\tilde{M}^{\s I'J'}_0
(\gamma_{\s I'J'}\gamma_{123})_{\gamma\delta}\,,
\label{final}
\end{eqnarray}
where $\tilde{M}_0^{\s IJ}$ and $\tilde{M}_0^{\s I'J'}$ are defined by  
\begin{eqnarray}
&&\tilde{M}^{\s IJ}_0 \;\equiv\; \int\!d^2\sigma\,
\left(\tilde{X}^{\s I}\tilde{P}^{\s J}-
\tilde{P}^{\s I}\tilde{X}^{\s J}
-\frac{1}{2}S\gamma^{\s IJ}\psi\right)\,,\,\,\nn\\
&&\tilde{M}^{\s I'J'}_0 \;\equiv\; \int\!d^2\sigma\,
\left(\tilde{X}^{\s I'}\tilde{P}^{\s J'}-
\tilde{P}^{\s I'}\tilde{X}^{\s J'}
-\frac{1}{2}S\gamma^{\s I'J'}\psi\right)\,.\,\,\nn
\end{eqnarray} 
Thus, we have proven that the zero-modes of 
the variables $X$ and $P$ in ``m'' (\ref{m}) 
have been subtracted and ``m'' can be described in terms of 
oscillation-modes in $X$ and $P$ only. 
That is, the matrix ``m'' represents
the excited modes of variables.  The first three terms 
in Eq.\,(\ref{final}) have the same forms as in the flat case, if 
2$\tilde{H}$ is replaced with the invariant mass $\mathcal{M}^2$.   
In the case of the
pp-wave, additional four terms proportional to $\mu$ appear. 
These terms impose the BPS conditions for the charges on
the additional extended objects only living on the pp-wave. 

We can find the BPS states by analyzing the rank of the 
supercharge matrix ``m''.  The 1/2 and 1/4 BPS cases will be 
considered in the following subsections.  When we consider the BPS states, 
the fermion parts are omitted in the discussion   
because we do not take the winding of fermion into account.

\subsection{1/2 BPS States}

The 1/2 BPS states can be described by 
the condition m=0, which means some constraints
\begin{eqnarray}
\label{first}
& &  2\tilde{H}-2\sum_{r,s=1}^9z_{rs}z^{rs}\;=\;0\,, \\ 
\label{second}
& & \tilde{z}_r \;=\; 0\,, \\
\label{third}
& & \tilde{U}_{\s JKI'J'} \;=\; \tilde{U}_{\s I'} \;=\; 0 \,,\\
& & \tilde{M}^{\s IJ}_0 \; =\; \tilde{M}^{\s I'J'}_0 \;=\; 0\,.
\end{eqnarray}
We neglected the term $z_{rs}z_{tu}\gamma^{rstu}$ 
by imposing proper conditions (for example, $z_{1r} \neq 0$ and 
otherwise is zero).
The first equation (\ref{first}) is the BPS condition 
indicating that the mass is proportional to its charge.  
The BPS condition also indicates that the oscillation-modes 
are absent. But there are modes 
linear to the worldvolume coordinates $\sigma^a$ ($a=1,2$) and 
$z_{rs} \neq 0$.
Then the second condition (\ref{second}) is automatically 
satisfied. The third condition should mean that the charge of 
the extended objects only living on the pp-wave. That is, 
this case corresponds to the (transverse) M2-brane 
in the flat space. It has been also shown in the flat space \cite{EMM2} that 
the longitudinal M2-brane is contained in 1/2 BPS states 
in the $SO(10,1)$ covariant way. Thus, it would be possible to 
say that the longitudinal M2-brane might
be contained in this case though it is not seen manifestly 
in our $SO(9)$ formalism.  
In conclusion, 1/2 BPS states are transverse and longitudinal 
M2-branes which are the same objects as the flat case. 

It is also possible to obtain the remaining unbroken 
supercharges under the 1/2 BPS
conditions. Note the supercharge matrix can be rewritten as 
\begin{eqnarray}
& & i\{ \dbinom{Q^{\uparrow}_{\al}}{Q^{\downarrow}_{\al}},\, (Q^{\uparrow}_{\beta}, Q^{\downarrow}_{\beta}) \}_{\rm\s DB} \;=\;
\begin{pmatrix}
{\bf -1}_{16}  & 0 \\
0 & {\rm m}
\end{pmatrix}
\,, \\ 
& & \dbinom{Q^{\uparrow}_{\al}}{Q^{\downarrow}_{\al}} \;\equiv\; 
\begin{pmatrix}
{\bf 1}_{16} & 0 \\
0 & \exp \left(\frac{\mu}{12}\gamma_{\s 123}\tau\right)
\end{pmatrix}
\begin{pmatrix}
{\bf 1}_{16} & 0 \\
- N_2 & {\bf 1}_{16} 
\end{pmatrix}
\dbinom{Q^-_{\al}}{Q^+_{\al}}\,.
\end{eqnarray}
We can easily find the unbroken supersymmetry and its charge is given by 
\begin{eqnarray}
 Q^{\downarrow} \;=\; - \e^{\frac{\mu}{12}\gamma_{\s 123}\tau}
\left(N_2Q^- - Q^+\right)\, .
\end{eqnarray}

\subsection{1/4 BPS States}

The 1/4 BPS conditions are that the rank of the matrix ``m'' 
is eight. 
In order to study the rank of ``m'', we decompose the 
$16 \times 16$ $SO(9)$ gamma matrices $\gamma^r$
$(r=1,\ldots,9)$ into $8\times 8$ gamma matrices as 
\begin{eqnarray}
 \gamma^{\tilde{r}} \;=\; 
\begin{pmatrix}
0 & \tilde{\gamma}^{\tilde{r}} \\ 
(\tilde{\gamma}^{\tilde{r}})^{\s T} & 0 
\end{pmatrix}\,~~(\tilde{r} = 1,\ldots, 8), 
\quad \gamma^9 \;=\; 
\begin{pmatrix}
{\bf 1}_{8} & 0 \\
0 & - {\bf 1}_{8}
\end{pmatrix}
\, ,
\end{eqnarray}
where $\tilde{\gamma}^{\tilde{r}}$'s and $\gamma^9$ 
are real and symmetric 
matrices with $8\times 8$ components
, and 
satisfy the commutation relations
\begin{eqnarray}
 \tilde{\gamma}^{\tilde{r}}(\tilde{\gamma}^{\tilde{s}})^{\s T} + \tilde{\gamma}^{\tilde{s}}(\tilde{\gamma}^{\tilde{r}})^{\s T} &=& 
2\del^{\tilde{r}\tilde{s}}\, , \nn \\
 (\tilde{\gamma}^{\tilde{r}})^{\s T}\tilde{\gamma}^{\tilde{s}} 
+ (\tilde{\gamma}^{\tilde{s}})^{\s T}\tilde{\gamma}^{\tilde{r}} 
&=& 2\del^{\tilde{r}\tilde{s}}\, . 
\end{eqnarray}
By the use of the above matrices, $\tilde{U}_{\s I'}(\gamma_{\s
I'}\gamma_{\s 123})$ and $\tilde{U}_{\s JKI'J'}(\gamma^{\s JKI'J'})$ 
can be rewritten as 
\begin{eqnarray}
& &  \sum_{{\s I'}=4}^9\tilde{U}_{\s I'}(\gamma_{\s
I'}\gamma_{\s 123})  \nn \\
&=& 
\begin{pmatrix}
\sum_{{\s \tilde{I}'}=4}^8
\tilde{U}_{\s \tilde{I}'}(\tilde{\gamma}^{\s \tilde{I}'})
(\tilde{\gamma}^1)^{\s T}\tilde{\gamma}^{2}(\tilde{\gamma}^3)^{\s T} & 
\tilde{U}_9 \tilde{\gamma}^1(\tilde{\gamma}^2)^{\s T}\tilde{\gamma}^3 \\
\label{u1} 
- \tilde{U}_9(\tilde{\gamma}^1)^{\s T}\tilde{\gamma}^2
(\tilde{\gamma}^3)^{\s T} & \sum_{{\s \tilde{I}'}=4}^8 
\tilde{U}_{\s \tilde{I}'} (\tilde{\gamma}^{\s \tilde{I}'})^{\s T} 
\tilde{\gamma}^1 (\tilde{\gamma}^2)^{\s T}\tilde{\gamma}^3 
\end{pmatrix}\, , \\
& & \sum_{{\s J,K}=1}^3\sum_{{\s I',J'}=4}^9
\tilde{U}_{\s JKI'J'}(\gamma^{\s JKI'J'}) \nn \\
&=& \sum_{{\s \tilde{J},\tilde{K}}=1}^3\sum_{{\s \tilde{I}',\tilde{J}'}=4}^8
\begin{pmatrix}
\tilde{U}_{\s \tilde{J}\tilde{K}\tilde{I}'\tilde{J}'}\tilde{\gamma}^{[ 
{\s \tilde{J}}}(\tilde{\gamma}^{\s \tilde{K}})^{\s T}
\tilde{\gamma}^{\s \tilde{I}'}(\tilde{\gamma}^{{\s \tilde{J}'}]})^{\s T}  & 0
\\
0 & 
\tilde{U}_{\s \tilde{J}\tilde{K}\tilde{I}'\tilde{J}'}(\tilde{\gamma}^{[{\s \tilde{J}}})^{\s T} \tilde{\gamma}^{\s \tilde{K}} (\tilde{\gamma}^{\s \tilde{I}'})^{\s T} \tilde{\gamma}^{{\s \tilde{J}'}]}
\end{pmatrix}
\nn \\
& & + \sum_{{\s \tilde{J},\tilde{K}}=1}^3\sum_{{\s \tilde{I}'}=4}^8
\begin{pmatrix}
0 &  - 2 
\tilde{U}_{\s \tilde{J}\tilde{K}\tilde{I}'9} \tilde{\gamma}^{[{\s \tilde{J}}}(\tilde{\gamma}^{\s \tilde{K}})^{\s T}\tilde{\gamma}^{{\s \tilde{I}'}]}  \\
2\tilde{U}_{\s \tilde{J}\tilde{K}\tilde{I}'9} 
(\tilde{\gamma}^{[ {\s \tilde{J}}})^{\s T}\tilde{\gamma}^{\s \tilde{K}}(\tilde{\gamma}^{{\s \tilde{I}'}]})^{\s T}  & 0
\end{pmatrix}
\,.
\label{u2}
\end{eqnarray}
We would like to consider the rank of the matrix ``m'', but 
the general analysis is more complicated and difficult. Here  
we shall present some special solutions concretely. 

To begin, we consider only $\mu$-dependent parts by letting
$\mu$-independent parts vanish.  
For simplicity, we impose further conditions $ 
\tilde{M}_0^{\s IJ}= \tilde{M}^{\s \tilde{I}'\tilde{J}'}_0 = 
\tilde{U}_{\s JKI'J'}= \tilde{U}_9 = 
0 $. 
Then the matrix ``m'' can be written as 
\begin{eqnarray}
{\rm m} &=& 
2\mu \sum_{{\s I'}=4}^9 \tilde{U}_{\s I'}(\gamma_{\s I'}\gamma_{\s 123})
- 2\cdot\frac{\mu}{6}\sum_{{\s I'}=4}^9\tilde{M}_0^{\s 9I'}
(\gamma_{\s 9 I'}\gamma_{\s 123})  \\
       &=& 2\mu
\begin{pmatrix}
\sum_{\s \tilde{I}'=4}^8 \left(\tilde{U}_{\s \tilde{I}'}-\frac{1}{6}
\tilde{M}_0^{9 {\s \tilde{I'}}}\right)\tilde{\gamma}^{\s \tilde{I}'}
(\tilde{\gamma}^1)^{\s T}\tilde{\gamma}^2(\tilde{\gamma}^3)^{\s T} & 0\\
0 & 
\sum_{\scriptstyle \tilde{I}'=4}^8 \left(\tilde{U}_{\s \tilde{I}'}+\frac{1}{6}
\tilde{M}_0^{9 {\s \tilde{I'}}}\right)\tilde{\gamma}^{\s \tilde{I}'}
(\tilde{\gamma}^1)^{\s T}\tilde{\gamma}^2(\tilde{\gamma}^3)^{\s T}
\end{pmatrix}\,. \nn
\end{eqnarray}
We can easily find the conditions that the rank of the matrix ``m'' is
eight, those are 
\begin{eqnarray}
\label{1/4BPS-1}
&& \tilde{U}_{\s\tilde{I}'} \;=\; +\frac{1}{6}\tilde{M}_0^{9\s\tilde{I}'} 
\qquad ({\scriptstyle \tilde{I}'} =4,\,5,\cdots ,8)\,,\\
\mbox{or}~~ && \tilde{U}_{\s\tilde{I}'} \;=\; 
-\frac{1}{6}\tilde{M}_0^{9\s\tilde{I}'} 
\qquad ({\scriptstyle \tilde{I}'}=4,\,5,\cdots ,8)\,.
\label{1/4BPS-2}
\end{eqnarray}
These conditions (\ref{1/4BPS-1}) and (\ref{1/4BPS-2}) 
show that the brane charge $\tilde{U}_{\s\tilde{I}'}$ equals to 
the angular momentum $\tilde{M}_0^{9 {\s \tilde{I}'}}$. 
These are 1/4 BPS solutions as special solutions 
of the general 1/4 BPS states. We may consider such solutions 
as rotating membranes which are 1/4 BPS. In fact, the superalgebra 
includes the angular momentum operators, which might be considered as 
the remnants of the $AdS_7 \times S^4$ or $AdS_4\times S^7$ backgrounds, 
and such rotating solutions can exist in the theory.   
Furthermore, (\ref{1/4BPS-1}) and 
(\ref{1/4BPS-2}) mean the ``stringy exclusion principle'' which is also 
a remnant in $AdS$ space physics. That is, configurations with
larger angular momenta than the given brane charges cannot exist.      
The central charges in the superalgebra may also indicate 
other extended objects only living on the pp-wave, which might be 
expected as a fuzzy membrane, a giant graviton \cite{Malda} or 
other 1/4 BPS states \cite{bak} coming from  Myers effects \cite{Myers}.   

In the above case, only the $\mu$-dependent parts have been studied, but 
we should note that such a situation can be also realized 
by taking the large $\mu$ limit 
without imposing certain conditions on $\mu$-independent parts.  
The large $\mu$ limit has been 
discussed in \cite{DSR} where the existence of such 1/4 BPS states is 
stated at least in this limit. 
Conversely speaking, in our considerations we could avoid the large $\mu$
limit by requiring $\mu$-independent parts 
to satisfy vanishing conditions. 

Moreover, we comment that it is possible to discuss the 1/4 BPS states 
in the same way as in the flat space \cite{EMM2}  
if we consider $\tilde{M}_0^{\s IJ} = \tilde{M}_0^{\s I'J'} = 
\tilde{U}_{\s I'} = \tilde{U}_{\s JKI'J'} =0$ (i.e., $\mu=0$).  
Also, it would be possible to apply 
the similar considerations for $\tilde{U}_{\s JKI'J'}$. 

Finally, the unbroken supercharges of the 1/4 BPS states are given by 
\begin{eqnarray}
 Q^{\downarrow (\mp)} \equiv \left(\frac{1 \mp \gamma^9}{2}
\right)Q^{\downarrow}\, ,
\end{eqnarray}
where the minus $(-)$ sign corresponds to the case in (\ref{1/4BPS-1}) and 
the plus $(+)$ one to (\ref{1/4BPS-2}), respectively.

\section{Conclusions and Discussions}

In this paper, we have obtained the BPS conditions of the
supermembrane on the pp-wave background from the viewpoint of the 
rank of the supercharge matrix by the use of triangular decomposition. 
In the pp-wave case the Hamiltonian and invariant mass 
contain more zero-modes than in the flat case, 
but we have seen that these zero-modes are spurious and 
play no roles in our considerations of the BPS states.
Moreover, we have in detail studied the BPS conditions for 
1/2 and 1/4 BPS states. 
The conditions for 1/2 BPS states are the same ones as in the flat space. 
For the 1/4 BPS states, 
we could not present general solutions but a special solution as an
example of the 1/4 BPS states, which is peculiar in the pp-wave case.

It is an interesting future work 
to investigate more general BPS mass formulae 
systematically.

\vspace*{0.5cm}
\noindent 
{\bf\large Acknowledgement}

The work of K.S. is supported in part by the Grant-in-Aid from the 
Ministry of Education, Science, Sports and Culture of Japan 
($\sharp$ 14740115).

\end{document}